# Chemically-adhesive particles form stronger and stiffer magnetorheological fluids

Abigail Rendos,[1] Daryl W. Yee,[2] Robert J. Macfarlane,[2] Keith A. Brown[1,3,4]

[1] Division of Materials Science and Engineering, Boston University, Boston, MA
[2] Department of Materials Science and Engineering, Massachusetts Institute of Technology, Cambridge, MA
[3] Department of Mechanical Engineering, Boston University, Boston, MA
[4] Department of Physics, Boston University, Boston, MA

E-mail: brownka@bu.edu

Magnetorheological fluids (MRF) are suspensions of magnetic particles that solidify in the presence of a magnetic field due to the particles forming chains along field lines. The magnetic forces between particles dominate the solidification process and determine the yield stress of the resulting solid. Here, we investigate how reversible chemical links between particles influence MRF behavior in terms of their yield stress and stiffness through rheological testing in flow and oscillation mode. Initially, we functionalize particles with phosphonate groups that are expected to link through hydrogen bonding and find that this MRF exhibits up to 40% higher yield stress and 100% higher stiffness than an MRF composed of unfunctionalized particles. To explain this change, we model the chemical attraction as an adhesion that supplements dipole-dipole interactions between particles. Interestingly, we find that the increase in yield stress is largest for dilute suspensions that are expected to solidify into isolated chains, while the proportional increase in yield stress is less for MRF with higher concentrations. This is explained by the higher concentration MRF forming a body-centered tetragonal (BCT) lattice in which interparticle adhesion forces are no longer aligned with the applied field. To explore the possibility of dynamically tuning interparticle interactions, we functionalize particles with polystyrene polymers with thymine terminal groups that will only exhibit interparticle hydrogen bonding in the presence of a small linking molecule, namely melamine. We find that MRF formed with these particles also exhibit up to a 40% increase in yield stress and ~100% increase in stiffness, but only in the presence of melamine. In addition to confirming the role of hydrogen bonding in increasing MRF stiffness and yield stress, these results highlight the possibility of dynamically tuning MRF performance using magnetic fields and chemical modifications.

1. Introduction

Magnetorheological fluids (MRF) are a type of smart fluid that reversibly solidify in the presence of a magnetic field.[1,2] MRF are widely used in a variety of applications including adaptive dampers, clutches, optical polishing, and more recently soft robotics.[3-8] In most applications, the yield stress $\tau_0$ of the solidified MRF is considered the main performance metric. As such, there has been extensive work exploring MRF formulations to increase $\tau_0$. Much of this work has focused on the composition, size, and shape of the magnetic particles[9,10] because the principal forces holding the solidified MRF together are from magnetic dipole-dipole interactions. Also of great interest is tuning the rheological properties of the fluid using additives such as thixotropic agents,[9] shear thickening agents,[11,12] nanoparticles,[13,14] and 2D materials[15,16] which can also influence the behavior of the MRF.

While tuning the particles or soluble additives have been important for increasing $\tau_0$ and reducing sedimentation, the interface between the particles and the medium has received comparatively little attention outside of its role in maintaining colloidal stability. In particular, when the particles are magnetically held together in a solid-like state, particle-particle contacts occur and the particles will chemically and sterically interact. The nature of this interaction and its effect on MRF performance has not been extensively studied but a few key requirements are clear:

For MRF to reversibly form in the presence of a magnetic field — or even rearrange upon shear-induced melting — it is crucial that the bonds be reversible. Indeed, reports of iron particles that were irreversibly polymerized into magnetic fibers showed a 30% decrease in $\tau_0$.[17, 18] In contrast, reversible adhesion, such as that provided by non-covalent interactions like hydrogen bonds, could provide an additional short ranged force to strengthen solidified particles (Fig. 1(a,b)). Indeed, early experiments using guar gum to coat magnetic particles resulted in a $\tau_0$ increase of up to 40% which the authors attributed to hydrogen bonding between particles mediated by the guar gum.[19] However, these results are complicated by the roughening effect of the guar gum coating. Specifically, the increase in $\tau_0$ was observed to be strongest in cases when the particles were functionalized non-uniformly with sparse amounts of guar gum. Further functionalization that led to more uniformly functionalized particles resulted in a smaller increase in $\tau_0$. Thus, further study is needed to isolate the effect of reversible chemical adhesion on MRF performance.

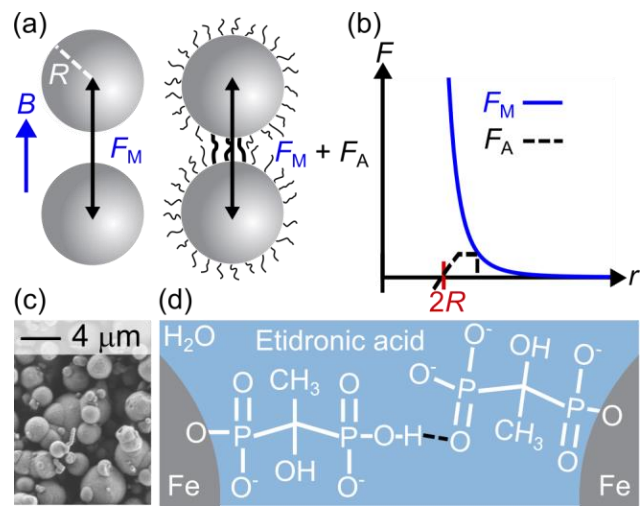

**Figure 1** (a) When magnetic particles with radius $R$ are aligned with $B$, they interact through the attractive magnetic dipole-dipole force $F_M$. Functionalized particles will have an additional attractive chemical adhesion force $F_A$. (b) $F_A$ is a short-ranged force that is maximum when particles are separated by $r$ and falls to zero as $r$ increases. $F_M$ scales as $r^{-4}$. (c) Scanning electron micrograph of carbonyl iron particles (d) Bonding scheme for phosphonate-functionalized magnetorheological fluid (pMRF) prepared through etidronic acid functionalization. Hydrogen bonds are indicated with dashed black lines.

Here, we explore the influence of adhesive surface functionalization on the performance of MRF through rheological testing. We initially focus on particles functionalized using etidronic acid, which is a small molecule that is expected to produce hydrogen bonding between phosphonate groups. Flow- and oscillation-mode experiments show that this phosphonate MRF (pMRF) exhibits up to a 40% increase in $\tau_0$ and a 100% increase in stiffness relative to uncoated, conventional MRF (cMRF). Interestingly, the relative $\tau_0$ enhancement decreased as $\varphi$ increased. We propose a model to explain the $\varphi$ dependence of $\tau_0$ attributing it to the differing structure of particle assembly from isolated chains at low $\varphi$ to a continuous body-centered tetragonal (BCT) structure at high $\varphi$ where the adhesive force $F_A$ is no longer aligned with the applied field. Building on this understanding, we hypothesize that it is possible to chemically switch between non-adhesive and adhesive particles using a second functionalization strategy in which the particles are coated with a thymine-presenting polymer that cross-links in the presence of melamine. Indeed, this functionalization strategy leads to a similar strengthening and stiffening as in the pMRF, but only in the presence of soluble melamine. The consistent results of these two systems along with the chemical tunability of the latter highlight novel routes for imparting stimuli responsiveness in smart fluids by chemically tuning particle interfaces.

## 2. Methods

*2.1 Synthesis of magnetorheological fluids*

*2.1.1 Phosphonate-functionalized MRF*

To synthesize MRF with chemically-adhesive particles, a method was adapted from Othmani *et al.*[20] to functionalize carbonyl iron microparticles (3-5 $\mu$m size, Skyspring Nanomaterials, 0990JH), shown in Fig. 1(c), with phosphonate groups that are predicted to adhere particles to one another through hydrogen bonding when they are in close contact, as depicted in Fig. 1(d). Specifically, a 1 mM etidronic acid solution was made by adding 17.2 µL of etidronic acid (60% aqueous solution, Sigma Aldrich, H6773) to 50 mL DI water which was neutralized to pH 7 using 10 mg sodium hydroxide (Sigma Aldrich, S8045). Then, 0.5 g of carbonyl iron was added and the vial was rotated on a mechanical spinner for 48 hours. After spinning, the particles were washed five times with DI water and then dried overnight.

Functionalized particles were then added to a 10 mM NaCl aqueous solution (Sigma Aldrich, S7653) at $\varphi$ = 2, 4, 8, or 14.2 vol% to make the pMRF. NaCl was included to screen any electrostatic repulsion that might arise from added charge from the etidronic acid on the particle surface.

*2.1.2 Conventional MRF*

To facilitate comparison with the pMRF, the particles for the cRMF were treated according to the same steps as above except without etidronic acid or sodium hydroxide being added to the solution prior to mixing in the mechanical stirrer. The iron particles were washed and dried in the same way and also added to the 10 mM NaCl solution at $\varphi$ = 2, 4, 8, or 14.2 vol%.

*2.1.3 Thymine functionalized MRF*

To prepare the thymine functionalized particles, we started by synthesizing a thymine-terminated polystyrene polymer with a phosphonate anchor group according to the procedure reported previously by Santos and Macfarlane.[21] In brief, we first synthesized the thymine-functionalized atom transfer radical polymerization (ATRP) initiator and the propargyl phosphate cyclohexamine salt. The thymine ATRP initiator was then used to synthesize a 10 kDa polystyrene polymer via ATRP. Following which, the alkyne functionalized phosphonate was added to the polymer via copper(I) catalyzed azide alkyne cycloaddition to yield the final thymine-functionalized polymer.

To graft the thymine-functionalized polymers onto the carbonyl iron particles, we first activated the surface of the particles by treating them with 0.5 M hydrochloric acid:[22] 2.0 g of particles were added to 30 mL of 0.5 M hydrochloric acid and then mechanically shaken for 40 min. The treated particles were washed eight times with DI water and then dried overnight under vacuum. To make each batch of thymine-functionalized particles, 250 mg of activated particles, 39 mg of thymine-functionalized polymer, and 1.5 mL of toluene were added to a 1.5 mL Eppendorf tube. The mixture was sonicated briefly to ensure that the polymers were fully dissolved in toluene before being mechanically shaken at 700 RPM at 70 ˚C for 15 hours. The thymine-functionalized particles were then washed six times with toluene and dried overnight under high vacuum. Finally, the functionalized particles were added to anisole at $\varphi$ = 2, 4, and 8 vol% to make the thymine-functionalized MRF (tMRF).

*2.1.4 Thymine-functionalized MRF with melamine*

In order to link the thymine-functionalized particles, melamine was used as a small linking molecule. First, a 99 mM melamine solution was made by mixing 25 mg of melamine (Sigma Aldrich, M2659) with 1 mL dimethylsulfoxide (Sigma Aldrich, D8418) and sonicating the solution for 10 min to fully dissolve the melamine. Then 1 mL of dimethylformamide (Sigma Aldrich, 72438) was added to the solution and vortexed until mixed. Directly before an experiment, 5 µL of the melamine solution was added to 995 µL of anisole (Sigma Aldrich, 123226) to form a 495 µM solution of melamine. Then, particles were added to the 495 µM melamine solution for $\varphi$ = 2, 4, and 8 vol% samples and immediately used in the rheometer where the field was applied in order to promote adhesion between assembled particles through hydrogen bonding, thus making tMRF + mel.

*2.2 Rheology*

All samples were sonicated for 2 min prior to rheological testing to break up any particle aggregates that resulted from the drying process. Then, 85 $\mu$L samples were immediately pipetted onto the rheometer (TA Instruments, DHR-1 with Magnetorheology Accessory) with a plate-plate geometry for testing and the rheometer gap height was set to 250 $\mu$m for all experiments. Prior to shearing the MRF, $B$ was applied to allow the particles to adhere when aligned. Specifically, magnetic field $B$ was applied in steps of 0.1 T from $B = 0.1$ to 0.8 T with 1 min between each step. Then, flow-mode rheology was conducted by varying the shearing strain rate $\dot{\gamma}$ from 20 to 200 s$^{-1}$ as the stress $\tau$ was measured and this was repeated for a given sample at each $B$, logarithmically spaced from 0.8 T decreasing down to 0.03 T. Samples were demagnetized between each $B$ to reduce the effects of residual magnetization. Each trial was repeated three times with a fresh sample.

Oscillation-mode rheology was conducted by loading the sample in the same manner as used in flow mode and then $B$ was increased from 0.1 T to 0.5 T in steps of 0.1 T, similar to the flow-mode measurements, to allow for adhesion of the particles prior to testing. Then, measurements were conducted at $B = 0.5$ T by setting a range of $\tau$ that encompassed the solid and liquid-like regions of the MRF at $B = 0.5$ T. This range varied for the different $\varphi$ tested. As $\tau$ was increased, the storage modulus $G'$ and loss modulus $G''$ were measured. Finally, each sample condition was repeated with a fresh sample for a total of three trials.

### 3. Results and Discussion

In order to determine the influence of adhesive particles on MRF performance, pMRF and cMRF were compared using flow-mode rheology. The measured $\tau$ exhibited by each fluid increased with $\varphi$, as shown in Fig. 2(a) for cMRF, and also increased dramatically with $B$. To quantify $\tau_0$, $\tau$ was measured *vs.* $\dot{\gamma}$ as shown in Fig. 2(b) for both the cMRF and pMRF. This data was fit to the Bingham plastic constitutive model which is frequently used to model MRF behavior,[2, 23-25]

$$\tau = \eta_{eff} \cdot \dot{\gamma} + \tau_0, \qquad (1)$$

where $\tau_0$ is the *y*-intercept of the curve and $\eta_{eff}$ is the effective viscosity. By repeating the measurements in Fig. 2(b) for logarithmically spaced $B$ between 0.03 and 0.8 T, $\tau_0$ was plotted *vs.* $B$ to compare both fluids. As shown in Fig. 2(c) for $\varphi = 2$ vol%, pMRF exhibited a higher $\tau_0$ than cMRF for all $B$ tested. For each condition, three trials were conducted and averaged. After repeating this process for all $\varphi$, the percent increase in $\tau_0$ due to the adhesive functionalization was calculated and plotted in Fig. 2(d). These data clearly show that the adhesive functionalized particles increased $\tau_0$ by up to 40%, but in a $\varphi$-dependent manner with sparse particle suspensions exhibiting greater strengthening. While the nature of the $\varphi$-dependence was not clear, the increase in $\tau_0$ validated the hypothesis that additional adhesive force strengthened the solidified MRF.

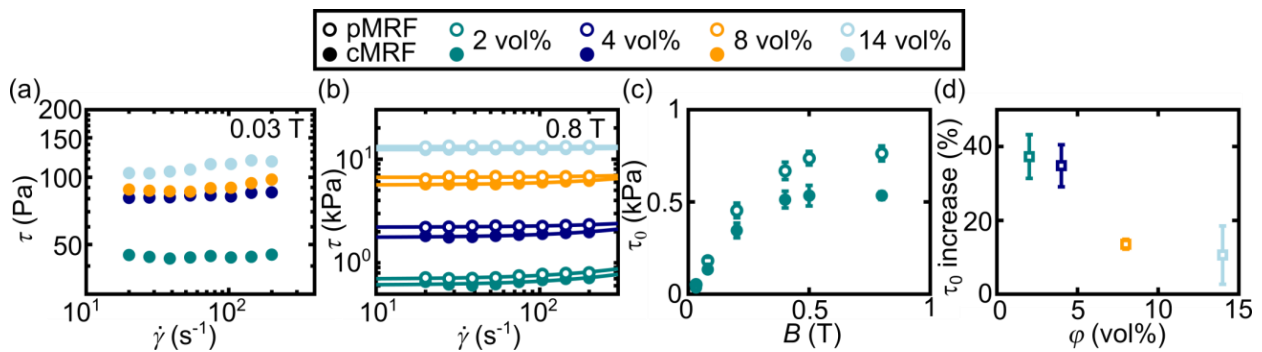

**Figure 2** (a) Representative plots of $\tau$ vs. $\dot{\gamma}$ for conventional MRF (cMRF) when $B = 0.03$ T for $\varphi = 2, 4, 8,$ and 14 vol% measured using flow-mode rheology. (b) Comparison of $\tau$ vs. $\dot{\gamma}$ between representative curves of pMRF and cMRF for all $\varphi$ at $B = 0.8$ T, the highest $B$ tested. Data was fit using the Bingham plastic model from Eq. 1 used to quantify the $\tau_0$. (c) $\tau_0$ vs. $B$ for $\varphi = 2$ vol% shows the increase in performance of pMRF compared to cMRF at all field strengths. (d) Percent increase in $\tau_0$ of pMRF relative to cMRF averaged over all $B$ for each $\varphi$.

Due to the added adhesive force for assembled particles, we hypothesized the solidified pMRF would exhibit higher stiffness than cMRF, which is another important performance metric for MRF applications.[26-28] In order to quantify the stiffness of the assembled particles, oscillation-mode rheology was used. In oscillation mode, shown in Fig. 3(a), the rheometer top plate oscillated back and forth in a $\tau$-controlled measurement where $\tau$ was applied and the strain $\gamma$ of the assembled particles was measured so that $G'$ and $G''$ of the material could be quantified. A representative plot of $G'$ and $G''$ for the cMRF is shown in Fig. 3(b). While high $\tau$ behaviour can be used to determine the onset of shear melting, the stiffness of the material is defined as the plateau $G'$ at low $\tau$. To compare the stiffness for the two fluids, $G'$ vs. $\tau$ was measured for $\varphi = 2$ and 8 vol%. In Fig 3(c), pMRF was observed to exhibit higher $G'$ than the cMRF. The stiffness was quantified by fitting the plot of $G'$ in loglog space to a power law with a *y*-offset which represented the extrapolated plateau $G_0$,

$$\log(G') = a * \log(\tau)^n + G_0 \qquad (2)$$

where *a* and *n* were additional fitting parameters. Interestingly, the pMRF plateau stiffness $G_0^p$ was about twice as large the cMRF plateau stiffness $G_0^c$ for both $\varphi$, as shown by their ratios in Fig. 3(d).

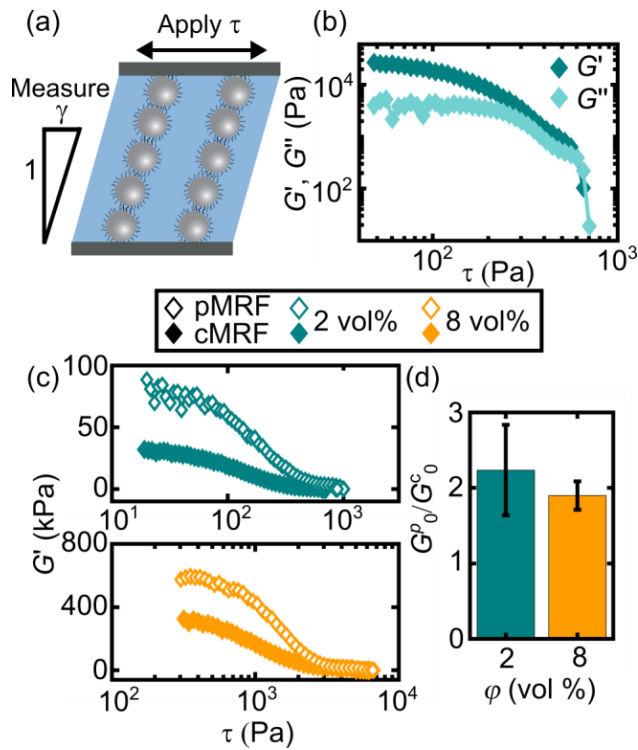

**Figure 3** (a) Scheme of oscillation-mode rheology. An oscillating $\tau$ was applied to the top plate and gradually increased while the strain $\gamma$ was measured. (b) An example plot of the data gathered from oscillation mode for the cMRF at $B = 0.5$ T where $G'$ and $G''$ were determined as a function of $\tau$. (c) Plots comparing $G'$ vs. $\tau$ for pMRF and cMRF at $\varphi = 2$ and 8 vol%. The pMRF exhibited significantly higher $G'$ plataeu at low $\tau$. (d) Plot of the plateau storage modulus for pMRF $G_0^p$ divided by the plateau storage modulus for cMRF $G_0^c$ for $\varphi = 2$ and 8 vol%. Error bars depict the error from the standard deviation of three trials of each fluid at each $\varphi$.

While the flow and oscillation-mode testing revealed that adhesive functionalization strengthened and stiffened the MRF, the strenghening was strongly $\varphi$ dependent with $\tau_0$ decreasing as $\varphi$ increased. In order to understand this dependence, we considered how the structure of the particle assembly changes as $\varphi$ increases. For low $\varphi$, MRF is understood to solidify through a series of isolated particle chains as depicted in Fig. 4(a) with a height determined by the rheometer gap. As a result, the magnetic dipoles that align with $B$ were also in alignment with $F_A$ so that both magnetic and adhesive contributions were acting in the same direction. However, at higher $\varphi$, MRF are known to form colums of BCT solids as shown in Fig. 4(b) due to interactions

ArXiv Pre-Print - Submitted 2/28/2022between neighboring chains causing isolated chains to combine.[29-31] Interestingly, MRF chains have been observed to interact and start forming columnar structures in both simulation and experiment at $\varphi > 5$ vol%[32, 33] which indicates that the transition from isolated chains to BCT columns is expected in range of $\varphi$ tested here. In this BCT structure, the interparticle contacts were no longer in the direction of the applied field, meaning that $F_A$ was also no longer aligned with the applied field. Therefore, as the chains were sheared, the $F_A$ between particles breaks at lower $\gamma$ than $F_M$ and contributed less to the strength of the assembled particles during failure. This likely explained why the strenthening associated with chemical adhesion decreased as $\varphi$ increased, whereas the increase in $G_0$ was similar between the two structures due to the stiffness being measured at very low $\gamma$ when the chains or columns are mainly intact.

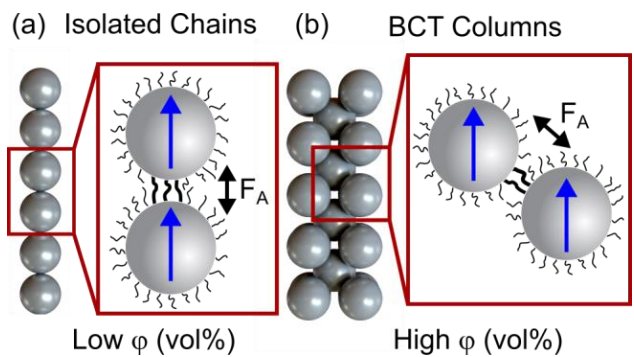

**Figure 4** (a) Scheme of isolated chains in an MRF at low $\varphi$. Within the isolated chains, the particles were aligned with the field due to the dipoles indicated in blue. $F_A$ was also aligned with the field, enabling a stronger contribution from $F_A$ to the strength of the chains. (b) Scheme of body-centered tetragonal (BCT) structure expected for high $\varphi$ MRF. Due to the structure, $F_A$ was not aligned with the dipoles causing the contribution from $F_A$ to be lower than for isolated chains.

Taking inspiration from the strengthening and stiffening observed for chemically-adhesive paticles, we hypothesized that we could leverage the dynamc tunability of chemistry to add additional stimuli responsiveness to MRF. Specifically, if the chemical adhesion of the particles could be modulated chemically, this could translate to large changes in the stiffness and strength of the solidified MRF. Thus, we synthesized tMRF, which was functionalized to present a thymine group. Without an additional linking molecule such as melamine, this is not expected to interact through hydrogen bonding. However, in the presence of melamine, these particles are expected to exhibit hydrogen bonding (Fig. 5(a)). To explore this idea, two systems were prepared, tMRF without melamine and tMRF with melamine (tMRF + mel). In contrast with the pMRF system, requiring a linking molecule for bonding endowed the system with more tunability and stimuli-responsiveness.

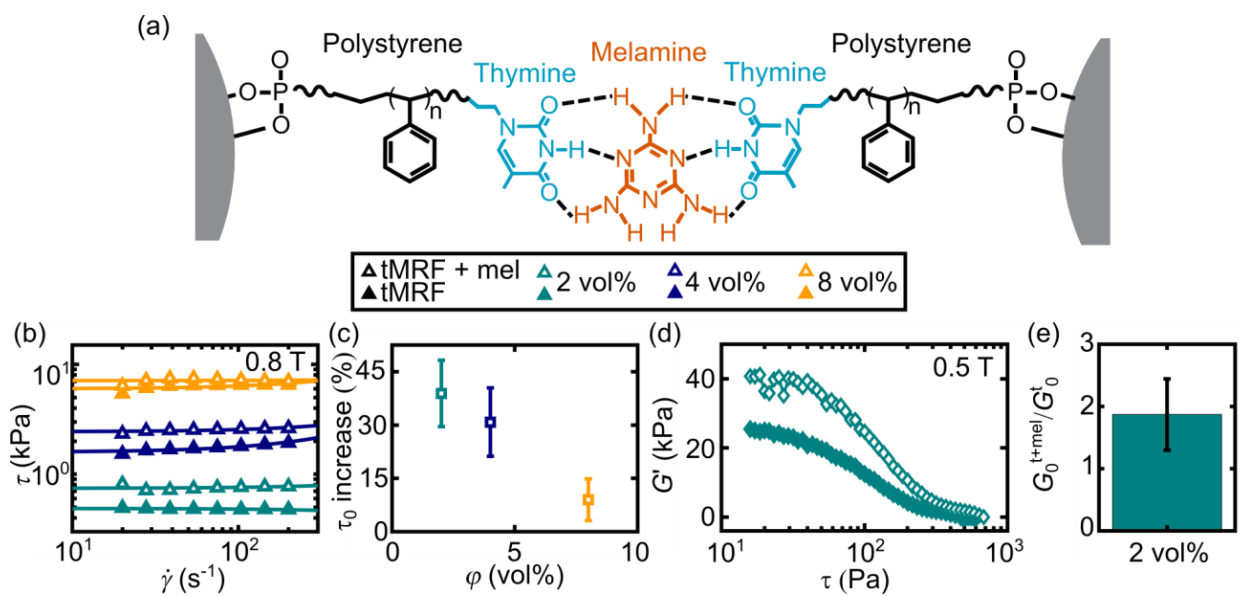

**Figure 5** (a) Bonding scheme for the thymine-functionalized MRF that was linked using the small molecule melamine (tMRF + mel). The thymine groups were attached to polystyrene-functionalized carbonyl iron particles. (b) Representative curves for $\tau$ vs. $\dot{\gamma}$ for the thymine-functionalized MRF (tMRF) and tMRF + mel at $B = 0.8$ T fit using Eq. 1. (c) Percent increase with error propogated from three trials taken for each fluid at $B = 0.8$ T for $\varphi = 2$, 4 and 8 vol%. (d) From oscillation-mode measurements, $G'$ vs. $\tau$ for $\varphi = 2$ vol% at $B = 0.5$ T. (e) Plot of the plateau storage modulus for tMRF + mel $G_0^{t+mel}$ divided by the plateau storage modulus for tMRF $G_0^t$ for $\varphi = 2$ vol%.

Flow-mode and oscillation-mode rheology were performed for the thymine-based system in order to contrast its behavior to the behavior of the pMRF. In Fig. 5(b), examples of $\tau$ vs. $\dot{\gamma}$ for the tMRF and tMRF + mel were included for $B = 0.8$ T and again fit to Eq. 1. In all cases, tMRF

+ mel exhibited higher $\tau_0$ than the tMRF and the percent increase is shown in Fig. 5(c) for $\varphi = 2$, 4, and 8 vol%. As with the pMRF, the percent increase in $\tau_0$ was about 40% for $\varphi = 2$ vol% and decreased with increasing $\varphi$, suggesting that the mechansism of strengthening was the same. Additionally, oscillation-mode measurements were conducted for $\varphi = 2$ vol%, as shown in Fig. 5(d), and were used to calculate the difference in stiffness between tMRF and tMRF+mel. The plateau stiffness $G_0^{t+mel}$ of tMRF with melamine was twice as large as the plateau stiffness $G_0^t$ of tMRF without melamine, in agreement with the comparison between pMRF and cMRF. Collectively, these results show that the tMRF+mel successfully maintained the same stiffening and strengthing as pMRF, but with a more flexible, dynamically linkable system.

## 4. Conclusion

Taken together, this work shows the potential for increasing performance metrics and introducing tunability of MRF through responsive surface coatings. Specifically, the performance of MRF with chemically-adhesive particles were compared to the performance of MRF formed by non-adhesive particles. In brief, chemically-adhesive particles drastically stiffened the solidified MRF in a manner independent of $\boldsymbol{\varphi}$. While these coatings also strengthened MRF by as much as 40%, this strengthening decreased as the $\boldsymbol{\varphi}$ increased. We attributed this decrease to the topological change between isolated chains and BCT solids as $\boldsymbol{\varphi}$ increased. It should be emphasized that the the fact that a small molecule can impart such a large change on the bulk performance of particles that are trillions of times their mass showcases the importance of tuning particle surfaces. In addition, this work highlights chemical tunability as an underutilized way with which to dynamically tune the properties of MRF and smart fluids more generally. Such chemical interactions could be utilized to locally tune the mechanics of MRF and even tune the temperature dependence of MRF properties.

**References**

1. Jolly, M. R.; Bender, J. W.; Carlson, J. D. *Journal of Intelligent Material Systems and Structures* **1999,** 10.
2. de Vicente, J.; Klingenberg, D. J.; Hidalgo-Alvarez, R. *Soft Matter* **2011,** 7, (8).
3. Çeşmeci, Ş.; Engin, T. *International Journal of Mechanical Sciences* **2010,** 52, (8), 1036-1046.
4. Wang, W.; Hua, X.; Wang, X.; Wu, J.; Sun, H.; Song, G. *Structural Control and Health Monitoring* **2019,** 26, (1).
5. Desrosiers, J. F.; Bigué, J. P. L.; Denninger, M.; Julió, G.; Plante, J. S.; Charron, F. *Journal of Physics: Conference Series* **2013,** 412.
6. McDonald, K.; Rendos, A.; Woodman, S.; Brown, K. A.; Ranzani, T. *Advanced Intelligent Systems* **2020,** 2, (11).
7. Abbott, J. J.; Ergeneman, O.; Kummer, M. P.; Hirt, A. M.; Nelson, B. J. *IEEE Transactions on Robotics* **2007,** 23, 1247-1252.
8. Kasai, T.; Jacobs, S. D.; Golini, D.; Hsu, Y.; Puchebner, B. E.; Strafford, D.; Prokhorov, I. V.; Fess, E. M.; Pietrowski, D.; Kordonski, W. I., Magnetorheological finishing: a deterministic process for optics manufacturing. In *International Conference on Optical Fabrication and Testing*, 1995; pp 372-382.
9. Ashtiani, M.; Hashemabadi, S. H.; Ghaffari, A. *Journal of Magnetism and Magnetic Materials* **2015,** 374, 716-730.
10. Kumar, J. S.; Paul, P. S.; Raghunathan, G.; Alex, D. G. *International Journal of Mechanical and Materials Engineering* **2019,** 14, (1).
11. Rendos, A.; Woodman, S.; McDonald, K.; Ranzani, T.; Brown, K. A. *Smart Materials and Structures* **2020,** 29, (7).
12. Zhang, X.; Li, W.; Gong, X. *Smart Materials and Structures* **2010,** 19, (12).
13. Wereley, N. M.; Chaudhuri, A.; Yoo, J. H.; John, S.; Kotha, S.; Suggs, A.; Radhakrishnan, R.; Love, B. J.; Sudarshan, T. S. *Journal of Intelligent Material Systems and Structures* **2006,** 17, (5), 393-401.
14. Jiang, W.; Zhang, Y.; Xuan, S.; Guo, C.; Gong, X. *Journal of Magnetism and Magnetic Materials* **2011,** 323, (24), 3246-3250.
15. Rendos, A.; Li, R.; Woodman, S.; Ling, X.; Brown, K. A. *Chem. Phys. Chem.* **2020,** 22, (5).
16. Zhang, W. L.; Choi, H. J. *Journal of Applied Physics* **2012,** 111, (7).
17. López-López, M. T.; Vertelov, G.; Bossis, G.; Kuzhir, P.; Durán, J. D. G. *Journal of Materials Chemistry* **2007,** 17, (36).
18. Kuzhir, P.; López-López, M. T.; Vertelov, G.; Pradille, C.; Bossis, G. *Rheologica Acta* **2007,** 47, (2), 179-187.
19. Wu, W. P.; Zhao, B. Y.; Wu, Q.; Chen, L. S.; Hu, K. A. *Smart Materials and Structures* **2006,** 15, (4), N94-N98.
20. Othmani, M.; Aissa, A.; Bac, C. G.; Rachdi, F.; Debbabi, M. *Applied Surface Science* **2013,** 274, 151-157.
21. Santos, P. J.; Macfarlane, R. J. *J Am Chem Soc* **2020,** 142, (3), 1170-1174.
22. Belyavskii, S. G.; Mingalyov, P. G.; Giulieri, F.; Combarrieau, R.; Lisichkin, G. V. *Protection of Metals* **2006,** 42, (3), 244-252.

23.     Weiss, K. D.; Duclos, T. G.; Carlson, J. D.; Chrzan, M. J.; Margida, A. J. *High Strength Magneto- and Electro-rheological Fluids*; SAE International: 1993.
24.     Ghaffari, A.; Hashemabadi, S. H.; Ashtiani, M. *Journal of Intelligent Material Systems and Structures* **2014,** 26, (8), 881-904.
25.     Claracq, J. r. m.; Sarrazin, J. r. m.; Montfort, J.-P. *Rheologica Acta* **2004,** 43, (1), 38-49.
26.     Li, W. H.; Du, H.; Chen, G.; Yeo, S. H.; Guo, N. *Rheologica Acta* **2002,** 42, (3), 280-286.
27.     Li, W. H.; Zhang, P. Q.; Gong, X. L.; Kosasih, P. B. *International Journal of Modern Physics B* **2005,** 19.
28.     Sun, S.; Tang, X.; Yang, J.; Ning, D.; Du, H.; Zhang, S.; Li, W. *IEEE Transactions on Industrial Informatics* **2019,** 15, (8), 4696-4708.
29.     Chin, B. D.; Park, J. H.; Kwon, M. H.; Park, O. O. *Rheologica Acta* **2001,** 40.
30.     Tang, X.; Conrad, H. *Journal of Physics D: Applied Physics* **2000,** 33.
31.     Tao, R. *Journal of Physics Condensed Matter* **2001,** 13, (50).
32.     Fernández-Toledano, J. C.; Rodríguez-López, J.; Shahrivar, K.; Hidalgo-Álvarez, R.; Elvira, L.; Montero de Espinosa, F.; de Vicente, J. *Journal of Rheology* **2014,** 58, (5), 1507-1534.
33.     Rodriguez-Lopez, J.; Castro, P.; Elvira, L.; Montero de Espinosa, F. *Ultrasonics* **2015,** 61, 10-4.